\documentclass[aps,pra,superscriptaddress,floatfix,showpacs,10pt, twocolumn]{revtex4}
\usepackage[dvips]{graphicx}
\usepackage{amsmath}

\begin{document}

\title{Scheme for implementing the Deutsch-Jozsa algorithm via atomic ensembles}
\author{Ping Dong}
\email{pingdong@ahu.edu.cn}
\author{Zheng-Yuan Xue}
\author{Ming Yang}
\author{Zhuo-Liang Cao}
\email{zlcao@ahu.edu.cn(Corresponding~Author)}

\affiliation{School of  Physics {\&} Material Science, Anhui
University, Hefei, 230039, People's Republic of China}


\begin{abstract}
We propose a physical scheme for implementing the Deutsch-Jozsa
algorithm using atomic ensembles and optical devices. The scheme has
inherent fault tolerance to the realistic noise and efficient
scaling with the number of ensembles for some entangled states
within the reach of current technology. It would be an important
step toward more complex quantum computation via atomic ensembles.
\end{abstract}
\maketitle

\section{Introduction}

Quantum computation is an enormously appealing task for
mathematicians, physical scientists and computer scientists because
of its potential to perform superfast quantum algorithms. Quantum
factoring \cite{Shor} and quantum search \cite{Grover} illustrate
the great theoretical promise of quantum computers. Quantum
entanglement is a striking feature of quantum mechanics, which can
be employed as a kind of ``quantum software'' to perform quantum
computation \cite{Gottesma}. A great deal of schemes for generating
entangled states have been proposed using cavity quantum
electrodynamics (QED) \cite{Zheng1,Guerra,Guo}, linear optics
\cite{Zhao} and so on \cite{Turchette,Gershenfeld,Jones}. As for
quantum computation, single-qubit manipulations and C-NOT gates
together can be served to realize any unitary operation on $n$
qubits. The C-NOT gate can be realized via many techniques
\cite{Pittman,Biswas,Wu}.

The Deutsch-Jozsa algorithm is a simple example of general quantum
algorithms, which can distinguish the function $f(x)$ between
constant and balanced \cite{Deutsch} on $2^{n}$ inputs $x$. The
values of the function $f(x)$ are either 0 or 1 for all possible
inputs. For the balanced function, the values of balanced function
are equal to 1 for half of all the possible inputs, and 0 for the
other half. The Deutsch-Jozsa algorithm has been realized in the
nuclear magnetic resonance system (NMR) \cite{Chuang}, ion trap
\cite{Gulde}, linear optical system \cite{Mohseni}, cavity QED
\cite{Zheng2} and atomic ensembles \cite{Dasgupta} theoretically and
experimentally.

Recently, many researchers have been paying their attentions to
atomic ensembles where the basic system is a large number of
identical atoms. A lot of interesting schemes for the generation of
quantum states and quantum information processing have been proposed
using atomic ensembles. For example, one can use atomic ensembles to
realize the scalable long-distance quantum communication
\cite{Duan1}, efficient generation of multipartite entanglement
states \cite{Duan2,Xue}, storage of quantum light \cite{Lukin,Liu}.
The schemes based on atomic ensembles have some special advantages
compared with the schemes of quantum information processing by the
control of single particles: (1) The schemes have inherent fault
tolerance and are robust to realistic noise and imperfections; (2)
Laser manipulation of atomic ensembles without separately addressing
the individual atoms is dominant easier than the coherent control of
single particles; (3) Atomic ensembles with suitable level structure
could have some kinds of collectively enhanced coupling to certain
optical mode due to the many-atom interference effects, which is
very important for all the recent schemes based on the atomic
ensembles. Here we  suggest the implementation of the Deutsch-Jozsa
algorithm via atomic ensembles. But our scheme is different from
Ref. \cite{Dasgupta} due to we chose different atomic configuration
and use different model. We utilize the collective enhancement of
atom ensembles other than considerate the interaction of atoms and
photon. Our scheme involves Raman-type laser manipulations, beam
splitters, and single-photon detections, the requirements of which
are well within  the current experimental technology. Obviously the
realization of Hadamard gate, C-NOT gate and equivalent Bell-basis
measurement is necessary to our ultimate aim. Therefore we will
introduce the scheme of these important implements in section II. In
section III, the process of implementing the Deutsch-Jozsa algorithm
is proposed in detail. The conclusions of the whole content are
given in the end section.

\section{Realization of C-NOT gate}

\begin{figure}[tbp]
\includegraphics[scale=0.3,angle=0]{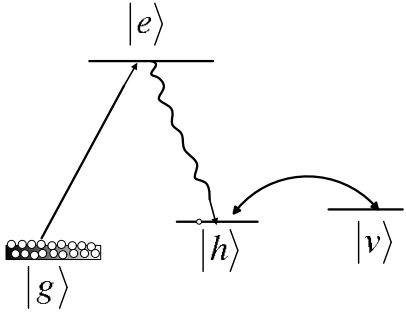}
\caption{The relevant atomic level structure of alkali metal atom.
$|g\rangle$ is the ground state, $|e\rangle$ is the excited state
and $|h\rangle$, $|v\rangle$ are two metastable states (Zeeman or
hyperfine sublevels) for storing a qubit of information. The
transition of $|e\rangle\rightarrow|h\rangle$ can emit a
forward-scattered Stokes photon that is co-propagating with the
laser pulse. The excitation in the mode $h$ can be transferred to
optical excitation by applying an anti-pump pulse.} \label{fig1}
\end{figure}

It has been shown that C-NOT gate can be realized theoretically with
the help of GHZ states, Bell-basis measures and Hadamard operations
from the proposal in Ref \cite{Gottesma}. In this section, we will
realize C-NOT gate using atomic ensembles with a large number of
identical alkali metal atoms as basic system. The scheme involves
Raman-type laser manipulations, beam splitters, and single-photon
detections. The relevant level structure of the alkali metal atoms
is shown in Fig. \ref{fig1}. For the three levels $|g\rangle$,
$|h\rangle $ and $|v\rangle $, two collective atomic operators can
be defined as
$S=(1/\sqrt{N})\Sigma_{i=1}^{N_{a}}|g\rangle_{i}\langle s|$, where
$s = h,v$, and $N_a \gg1$ is the total number of atoms. The three
levels $|g \rangle$, $|h\rangle$ and $|v\rangle$ can be coupled via
a Raman process. The atoms are initially prepared in the ground
state $|g \rangle$ using optical pump. $S$ is similar to independent
bosonic mode operators if only the atoms are all remain in ground
state $|g \rangle$. The states of the atomic ensemble can be express
as $|S \rangle = s^{+}|vac\rangle$ ($s = h,v)$ after the emission of
the single Stokes photon in a forward direction, where $|vac\rangle
$ denotes the ground state of atomic ensembles and $|vac \rangle
\equiv \otimes _{i}|g\rangle _{i}$. Long time coherence has been
demonstrated experimentally both in a room-temperature dilute atomic
gas \cite{Phillips} and in a sample of cold trapped atoms
\cite{Liu}. The above character of atoms is useful for the
generation of entanglement between atomic ensembles \cite{Duan2,
Xue}. Single-qubit operations with high precision between the two
atomic states $|S\rangle = s^ {+}|vac\rangle (s = h,v)$ can be
completed by simply shinning Raman pulses or radio-frequency pulses
on all the atoms. For instance, we can obtain $h^ {+} |vac \rangle
\rightarrow(h^ {+} + v^{+})|vac\rangle/\sqrt{2}$ and $v^ {+} |vac
\rangle \rightarrow(h^ {+} - v^{+})|vac\rangle/\sqrt{2}$ by choosing
appropriate length of the pump and anti-pump pulse. It is equivalent
to realize a Hadamard gate, which is very useful in the below
scheme.

\begin{figure}[tbp]
\includegraphics[scale=0.3,angle=90]{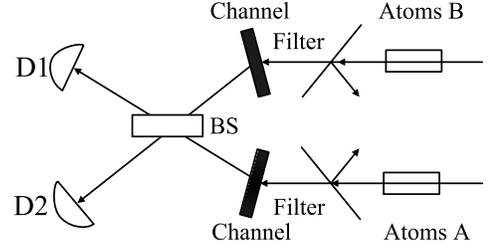}
\caption{Setup of realizing Bell-basis measurement. The two atomic
ensembles A and B are pencil-shaped, which are illuminated by the
synchronized laser pulses. The forward-scattered anti-Stokes are
collected and coupled to optical channel (fiber) after the filter.
BS is a 50/50 beam splitter, and the outputs detected by two
single-photon detectors $D$1 and $D$2.} \label{fig2}
\end{figure}

Bell-basis measurement is very important in the process of realizing
C-NOT gate, quantum information processing and quantum computation.
Therefore it is necessary to introduce the realization of the
Bell-basis measurement based on atomic ensembles in our paper. The
four Bell states in the system are $|\phi\rangle _{AB}^\pm = (
h_{A}^ {+} h_{B}^{+} \pm v_{A}^ {+} v_{B}^ {+})|vac\rangle
_{AB}/\sqrt{2}$
 and $|\varphi\rangle _{AB}^\pm = ( h_{A}^
{+} v_{B}^{+} \pm v_{A}^ {+}h_{B}^ {+})|vac\rangle _{AB}/\sqrt{2}$.
We can use the setup to achieve the task, as shown in Fig.
\ref{fig2}. Firstly, we apply anti-pump laser pulses to the two
atomic ensembles $ A $ and $ B$ to transfer their $h$ excitations to
optical excitations, and detect the anti-Stokes photons by detectors
$D$1 and $D$2. In the case of only detector $D$1 (or $D$2) clicks.
We will apply single-qubit rotations to both ensembles to rotate
their $v$ modes to $h$ modes by shinning $\pi$ length Raman pulses
or radio-frequency pulses on the two ensembles $A$ and $B$. Then we
apply anti-pump laser pulses to the two atomic ensembles $A$ and
$B$again, and detect anti-Stokes photons by $D$1 and $D$2. Now,
there are two different results of detection: (1) If detector $D$1
(or $D$2) clicks (i.e. only one detector clicks in the two
detections), post-select the cases that each ensemble has only one
excitation, atomic ensembles $A$ and $B$ are projected to
$|\varphi\rangle _{AB}^+= ( h_{A}^ {+} v_{B}^{+}+ v_{A}^ {+}h_{B}^
{+})|vac\rangle _{AB}/\sqrt{2}$; (2)If $D$2 (or $D$1) clicks
(\emph{i.e}. detectors $D$1 and $D$2 click respectively in the two
detections), post-select the cases that each ensemble has only one
excitation, atomic ensembles $A$ and $B$ are projected to
$|\varphi\rangle _{AB}^-= ( h_{A}^ {+} v_{B}^{+}- v_{A}^ {+}h_{B}^
{+})|vac\rangle _{AB}/\sqrt{2}$. Obviously, if we add single-qubit
rotations in the above process and repeat the above process of (1),
we can realize the projection of $|\phi\rangle _{AB}^\pm = ( h_{A}^
{+} h_{B}^{+} \pm v_{A}^ {+}v_{B}^ {+})|vac\rangle _{AB}/\sqrt{2}$
by post-selecting sense.

In order to realize C-NOT gate, we prepare two GHZ states, which
have been made in Ref \cite{Duan2} with atomic ensembles 1, 2, 3, 4,
5 and 6

\begin{subequations}
\begin{equation}
|\phi\rangle _{123} =(h_{1}^ {+} h_{2}^{+} h_{3}^ {+} + v_{1}^{+}
v_{2}^ {+} v_{3}^ {+})|vac\rangle_{123}/{\sqrt 2 },
\end{equation}
\begin{equation}
|\varphi\rangle _{456} = (h_{4}^ {+} h_{5}^{ +} h_{6}^ {+} + v_{4}^
{+} v_{5}^{+} v_{6}^{+})|vac\rangle_{456}/{\sqrt 2 },
\end{equation}
\end{subequations}
and prepare two atomic ensembles 7 and 8, which are in $|\phi
\rangle _{7} = ( h_{7}^{+} + v_{7}^{+})|vac\rangle_{7}$
 and $|\phi\rangle _{8} =(h_{8}^{+} - v_{8}^ {+} )|vac
\rangle _{8}$ by single-qubit rotations. The C-NOT gate has $|\phi
\rangle _{7}$ as its control, and $|\phi\rangle_{8}$ as its target.
At first, we apply Hadamard transformations on atomic ensembles 1, 2
and 3 by single-qubit operations, respectively, and then make a
Bell-basis measurement on atomic ensembles 3 and 4 using the setup
in Fig. \ref{fig2}. Here, the state of the $|\varphi\rangle
_{123456}$ collapses to one of the following four unnormalized
states
\begin{subequations}
\begin{eqnarray}
 |\phi\rangle_{1256}&=&[(h_{1}^{+}h_{2}^{+} +
v_{1}^{+}v_{2}^ {+})h_{5}^{+} h_{6}^{+}\nonumber\\
&\pm&(h_{1}^{+}v_{2}^{+}+v_{1}^{+}h_{2}^{+})v_{5}^{+}v_{6}^{+}]|vac\rangle_{1256},
\end{eqnarray}
\begin{eqnarray}
|\varphi\rangle_{1256}&=&[(h_{1}^{+}h_{2}^{+}+v_{1}^{+}v_{2}^{+})v_{5}^{+}v_{6}^{+}\nonumber\\
&\pm&(h_{1}^{+}v_{2}^{+}+v_{1}^{+}h_{2}^{+})h_{5}^{+}h_{6}^{+}]|vac\rangle_{1256}.
\end{eqnarray}
\end{subequations}
$|\phi\rangle_{1256}$ and $|\varphi\rangle _{1256}$ are the results
of projecting to $|\phi\rangle _{34}^\pm $ and $|\varphi \rangle
_{34}^\pm $, respectively. They can unify as
$|\chi\rangle_{1256}=[(h_{1}^{+}h_{2}^{+} + v_{1}^{+}v_{2}^
{+})h_{5}^{+} h_{6}^{+}+(h_{1}^{+}v_{2}^{+}
+v_{1}^{+}h_{2}^{+})v_{5}^{+}v_{6}^{+}]|vac\rangle_{1256}$ with the
help of simple single-qubit operations.

In succession, we make a Bell-basis measurement on atomic ensembles 1 and 8
as the above techniques. The state of atomic collective 1, 2, 5, 6 and 8
collapses to one of the following two states
\begin{subequations}
\begin{eqnarray}
\label{256a} |\phi\rangle_{256}&=&[(h_{2}^ {+}h_{5}^{+}h_{6}^{+}+
v_{2}^{+}v_{5}^ {+}v_{6}^{+})\nonumber\\
&+&(v_{2}^{+}h_{5}^{+}h_{6}^{+}+h_{2}^{+}v_{5}^{+}v_{6}^{+})]|vac\rangle_{256},
\end{eqnarray}
\begin{eqnarray}
\label{256b} |\varphi\rangle_{256}&=&[(h_{2}^ {+}h_{5}^{+}h_{6}^{+}
+ v_{2}^{+}v_{5}^ {+}v_{6}^{+})\nonumber\\
&-&(v_{2}^{+}h_{5}^{+}
h_{6}^{+}+h_{2}^{+}v_{5}^{+}v_{6}^{+})]|vac\rangle_{256},
\end{eqnarray}
\end{subequations}
where Eq. (\ref{256a}) is the result of projecting to $|\phi \rangle
_{18}^ {+}$ and $|\varphi \rangle_{18}^ + $, and Eq. (\ref{256b}) is
the result of projecting to $|\phi\rangle_{18}^{-} $ and $|\varphi
\rangle_{18}^{-}$. Apply single rotations to Eq. (\ref{256b}), leads
the state of atomic collective 2, 5 and 6 to Eq. (\ref{256a}). We
make a Bell-basis measurement on atomic ensembles 6 and 7, the state
of atomic collective 2 and 5 collapses to the one of the following
states
\begin{subequations}
\begin{equation}
\label{25a}|\phi\rangle_{25}=(h_2^{+} - v_2^{+})(h_5^{+}
-v_5^{+})|vac\rangle_{25}/2,
\end{equation}
\begin{equation}
\label{25b}|\varphi\rangle_{25}=(h_2^{+} - v_2^{+})(h_5^{+}
+v_5^{+})|vac\rangle_{25}/2.
\end{equation}
\end{subequations}
where Eq. (\ref{25a}) corresponds to the measurement results of
$|\phi \rangle _{67}^{+}$ and $|\varphi\rangle_{67}^{+}$, and Eq.
(\ref{25b}) corresponds to $|\phi\rangle_{67}^{-} $ and $|\varphi
\rangle_{67}^{-}$. We can make state (\ref{25b}) transform to state
(\ref{25a}) by single-qubit rotations. Obviously here, the C-NOT
gate has been realized and the state have mapped on atomic
collective 2 and 5. Therefore if only we apply Hadamard
transformations, Bell-basis measurements and single rotations on
atomic ensembles, we are able to realize C-NOT gate perfectly.

\section{Realization of the Deutsch-Jozsa algorithm}

The Realization of Deutsch-Jozsa algorithm is an important step
toward more complex quantum computation. Classically, if we want to
distinguish $f(x)$ between constant and balanced function on $2^{n}$
inputs, we will need $2^{n}/2+1$ queries to unambiguously determine
whether the function is balanced. While for the Deutsch-Jozsa
algorithm, we will need only one query. Here, we consider the
two-qubit Deutsch-Jozsa algorithm. The input query qubit is prepared
in $(|0\rangle_{i} + |1\rangle_{i})/\sqrt{2}$ and the auxiliary
working qubit is prepared in
$(|0\rangle_{a}-|1\rangle_{a})/\sqrt{2}$, so the state of the whole
system is $(|0\rangle_{i} +
|1\rangle_{i})(|0\rangle_{a}-|1\rangle_{a})/2$. While the function
$f(x)$ is characterized by unitary mapping transformation $U_{f}$,
and $|x,y\rangle \rightarrow|x,y \oplus f(x)\rangle$ where $\oplus$
indicates addition modulo 2. The unitary transformation $U_{f}$ on
the system leads the initially state to
\begin{equation}
\label{5}
[(-1)^{f(0)}|0\rangle_{i}+(-1)^{f(1)}|1\rangle_{i}](|0\rangle_{a}-|1\rangle_{a})/2.
\end{equation}
There are four possible transformations to the $U_{f}$:
\begin{enumerate}
    \item for $U_{f1}$, $f(0)=f(1)=0$;
    \item for $U_{f2}$, $f(0)=f(1)=1$;
    \item for $U_{f3}$, $f(0)=0$ and $f(1) = 1$;
    \item for $U_{f4}$, $f(0)=1$ and $f(1)=0$. \\
\end{enumerate}
After a Hadamard transformation on the input query qubit, the state
of query qubit becomes $|f(0)\oplus f(1)\rangle$. If the function
$f(x)$ is constant, the state of query qubit becomes $|0 \rangle_{i}
$. Otherwise it becomes $|1\rangle_{i}$.

Now, we realize the two-qubit Deutsch-Jozsa algorithm via atomic
ensembles and linear optical elements. At first, we prepare two
atomic ensembles 9 and 10, which are initially in the states $|\phi
\rangle_{9} = h_{9}^{+}|vac\rangle_{9}$ and $|\phi\rangle_{10} =
h_{10}^{+}|vac\rangle_{10}$, respectively. After two single-qubit
operations by controlling Raman pulses with the appropriate length,
we can obtain the state of the atomic collective 9 and 10 is
\begin{equation}
|\phi\rangle_{910} = (h_{9}^{+}+
v_{9}^{+})(h_{10}^{+}-v_{10}^{+})|vac\rangle_{910}/2,
\end{equation}
which can be rewritten as
\begin{equation}
|\phi\rangle_{910} =(|0\rangle_{9} + |1\rangle_{9})(|0 \rangle_{10}
- |1\rangle_{10})/2,
\end{equation}
where $|0\rangle_{9} = h_{9}^{+}|vac\rangle_{9}$, $|1\rangle_{9} =
v_{9}^{+}|vac\rangle_{9} $, $|0\rangle_{10}=h_{10}^ {+}|vac\rangle
_{10}$ and $|1\rangle_{10} = v_{10}^{+}|vac\rangle_{10}$. So the
collective state has the same form as that of input and auxiliary
working qubit in the Deutsch-Jozsa algorithm.

For the case of performing $U_{f1} $, we take no operation on the
system and the state remains in the state $|\phi\rangle _{910}$.

For the case of performing $U_{f2} $, we can make a C-NOT
transformation on the two collective 9 and 10 (collective 9 as
control bit, collective 10 as target bit), which can be achieved
using the scheme in section II. This leads to
\begin{equation}
\label{910}
|\phi\rangle_{910}=[|0\rangle_{9}(|0\rangle_{10}-|1\rangle_{10})+|1\rangle_{9}(|0\rangle_{10}\oplus1-|1\rangle_{10}\oplus1)]/2,
\end{equation}
Then we perform single-qubit operation on ensemble 9 using a
Raman-type pulse
\begin{subequations}
\label{9102}
\begin{equation}
|0\rangle_{9}\rightarrow|1\rangle_{9} ,
\end{equation}
\begin{equation}
|1\rangle_{9}\rightarrow-|0\rangle_{9} .
\end{equation}
\end{subequations}
and a C-NOT transformation on the two collective 9, 10 lead Eq.
(\ref{910}) to
\begin{eqnarray}
\label{9103}
|\phi\rangle_{910}&=&[|1\rangle_{9}(|0\rangle_{10}\oplus1-|1\rangle_{10}\oplus1)\nonumber\\
&-&|0\rangle_{9}(|0\rangle_{10}\oplus1-|1\rangle_{10}\oplus1)]/2,
\end{eqnarray}
then another single-qubit operation
\begin{subequations}
\label{9104}
\begin{equation}
|0\rangle_{9}\rightarrow-|1\rangle_{9} ,
\end{equation}
\begin{equation}
|1\rangle_{9}\rightarrow|0\rangle_{9} .
\end{equation}
\end{subequations}
lead the state of system to
\begin{equation}
|\phi\rangle_{910} = (-|0\rangle_{9}-|1\rangle_{9})(|0\rangle_{10}-
|1\rangle_{10})/2.
\end{equation}

For the case of performing $U_{f3} $,   we only need a C-NOT
transformation on the two collective as  Eq. (\ref{9103}), then the
system becomes
\begin{equation}
|\phi\rangle_{910}=(|0\rangle_{9}-|1\rangle_{9})(|0\rangle_{10}-|1\rangle_{10})/2.
\end{equation}

For the case of performing $U_{f4} $, we first perform a
single-qubit transformation of Eq. (\ref{9102}), and then we perform
a C-NOT operation of Eq. (\ref{910}). Finally, we perform a
single-qubit transformation of Eq. (\ref{9104}). This leads to
\begin{equation}
|\phi\rangle_{910}=(-|0\rangle_{9}+ |1\rangle_{9})(|0\rangle_{10} -
|1\rangle_{10})/2
\end{equation}

Obviously we can realize the unitary transformation $U_f $ by the
above method, the state of system can be expressed as
\begin{equation}
|\phi\rangle_{910}=[(-1)^{f(0)}|0\rangle_{9}+(-1)^{f(1)}|1\rangle_{9}](|0\rangle_{10}-|1\rangle_{10})/2.
\end{equation}
This state of the two collective 9 and 10 has the same form as the
Eq. (\ref{5}). Then we perform a Hadamard transformation on the
collective 9 by a single-qubit operation, which has been mentioned
in section II. Finally we detect the state of collective 9. If the
state is $h_{9}^{+}|vac\rangle_{9}$, the function $f(x)$ is
constant. Otherwise, $f(x)$ is balanced function. Now, we have
realized the Deutsch-Jozsa algorithm perfectly. Obviously a
measurement is sufficient to distinguish $f(x)$ between constant and
balanced function.

\section{Conclusions}

In summary, we present a physical scheme for implementing the
Deutsch-Jozsa algorithm with atomic ensembles, and the basic system
is a large number of identical alkali metal atoms. The Bell-basis
measurement and C-NOT gate are very important in quantum information
processing and quantum computation, thus the schemes with atomic
ensembles for realizing them are proposed firstly. Then we apply
Bell-basis measurements and C-NOT transformations into the process
of realizing Deutsch-Jozsa algorithm. The above schemes have some
special advantages compared with the schemes of quantum information
process by the control of single particles. For instance, it has
inherent fault tolerance to the realistic noise and imperfections of
entanglement generation, collectively enhanced coupling to light
because of special atomic configuration and many-atom interference
effects. Although the Deutsch-Jozsa algorithm is a simple example of
general quantum algorithms, it is very necessary to achieve quantum
computer. Further more, it would be an important step toward more
complex quantum computation. The scheme involves laser manipulations
of atomic ensembles, beam splitters, and single-photon detections
with moderate efficiencies, which are all within the current
experimental technology.

\begin{acknowledgments}

This work is supported by the Natural Science Foundation of the
Education Department of Anhui Province under Grant No: 2004kj005zd
and Anhui Provincial Natural Science Foundation under Grant No:
03042401 and the Talent Foundation of Anhui University.
\end{acknowledgments}

\end{document}